\documentclass[conference]{IEEEtran}
\usepackage{cite}
\usepackage{amsmath,amssymb,amsfonts}
\usepackage{algorithm}
\usepackage{algpseudocode}
\usepackage{graphicx}
\usepackage{textcomp}
\usepackage[dvipsnames]{xcolor}
\usepackage[hyphens]{url}
\usepackage{fancyhdr}
\usepackage{array}
\usepackage{multirow}
\usepackage{booktabs}
\usepackage{adjustbox}
\usepackage{wrapfig}
\usepackage{tikz}
\usepackage{bm}
\usepackage{lipsum} 
\usepackage{xspace}
\usepackage{pifont}
\usepackage{tablefootnote}
\usepackage[bookmarks=true,breaklinks=true,letterpaper=true,colorlinks,citecolor=blue,linkcolor=blue,urlcolor=blue]{hyperref}

\newcommand{\sym}[1]{{#1}}

\newcommand{\name}{\sym{LoAS}\xspace}

\newcommand{\minisection}[1]{\vspace{0.05in}\noindent {\bf #1}}

\newcommand*\circled[1]{\tikz[baseline=(char.base)]{
            \node[shape=circle,fill,inner sep=0.8pt] (char) {\textcolor{white}{#1}};}}
            
\newcommand*\colourcheck[1]{%
  \expandafter\newcommand\csname #1check\endcsname{\textcolor{#1}{\ding{52}}}%
}

\newcommand*\colourcross[1]{%
  \expandafter\newcommand\csname #1cross\endcsname{\textcolor{#1}{\ding{56}}}%
}

\colourcheck{blue}
\colourcheck{teal}
\colourcross{red}

\title{LoAS: Fully Temporal-Parallel Dataflow for Dual-Sparse Spiking Neural Networks}
\begin{document}

\author{\IEEEauthorblockN{Ruokai Yin}
\IEEEauthorblockA{
Yale University\\
New Haven, USA\\
ruokai.yin@yale.edu}
\and
\IEEEauthorblockN{Youngeun Kim}
\IEEEauthorblockA{
Yale University\\
New Haven, USA\\
youngeun.kim@yale.edu}
\and
\IEEEauthorblockN{Di Wu}
\IEEEauthorblockA{
University of Central Florida\\
Orlando, USA\\
di.wu@ucf.edu}
\and
\IEEEauthorblockN{Priyadarshini Panda}
\IEEEauthorblockA{
Yale University\\
New Haven, USA\\
priya.panda@yale.edu}
}

\maketitle
\pagestyle{plain}

%%%%%% -- PAPER CONTENT STARTS-- %%%%%%%%

\begin{abstract}
Spiking Neural Networks (SNNs) have gained significant research attention in the last decade due to their potential to drive resource-constrained edge devices.
Though existing SNN accelerators offer high efficiency in processing sparse spikes with dense weights, opportunities are less explored in SNNs with sparse weights, i.e., dual-sparsity. 
In this work, we study the acceleration of dual-sparse SNNs, focusing on their core operation, sparse-matrix-sparse-matrix multiplication (spMspM). 
We observe that naively running a dual-sparse SNN on existing spMspM accelerators designed for dual-sparse Artificial Neural Networks (ANNs) exhibits sub-optimal efficiency.
The main challenge is that processing timesteps, a natural property of SNNs, introduces an extra loop to ANN spMspM, leading to longer latency and more memory traffic.
To address the problem, we propose a fully temporal-parallel (FTP) dataflow, which minimizes both data movement across timesteps and the end-to-end latency of dual-sparse SNNs.
To maximize the efficiency of FTP dataflow, 
we propose an FTP-friendly spike compression mechanism that efficiently compresses single-bit spikes and ensures contiguous memory access. 
We further propose an FTP-friendly inner-join circuit that can lower the cost of the expensive prefix-sum circuits with almost no throughput penalty. 
All the above techniques for FTP dataflow are encapsulated in \name, a \underline{Lo}w-latency inference \underline{A}ccelerator for dual-sparse \underline{S}NNs. 
With FTP dataflow, compression, and inner-join, running dual-sparse SNN workloads on \name demonstrates significant speedup (up to $8.51\times$) and energy reduction (up to $3.68\times$) compared to running it on prior dual-sparse accelerators.

\end{abstract}

\section{Introduction}

Spiking Neural Networks (SNNs) have attracted considerable interest as potential energy-efficient substitutes for Artificial Neural Networks (ANNs)~\cite{roadmap_snn,spike_nature,fang2023spikingjelly}.
Inspired by the biological neuron, SNNs leverage highly sparse unary-coded (\{0,1\}) spikes to compute and communicate information~\cite{wu2020ugemm}. 
Thus, running SNNs on hardware significantly reduces computation and data movement, making it suitable for edge computing.
Therefore, SNNs have been widely used in computer vision tasks, such as image classification~\cite{sengupta2019going,wu2018spatio}, optical flow estimation~\cite{lee2020spike}, semantic segmentation~\cite{kim2022beyond}, and object detection~\cite{kim2020spiking}.

\minisection{Opportunity.} As the need for edge devices with limited memory capacity increases, recent research on SNNs highlights the significance of dual-sparse (both spikes and weights are sparse), which can be achieved by neural pruning techniques~\cite{kim2022exploring,roadmap_snn}. 
Pruning the weight connections of SNNs has been explored during both training~\cite{chen2021pruning,shi2019soft} and inference~\cite{neftci2016stochastic}.
Certain works have managed to achieve approximately 98\% weight sparsity and 90\% spike sparsity~\cite{kim2022exploring}, leveraging the lottery ticket hypothesis~\cite{frankle2018lottery}.
These works have outlined the potential of dual-sparse SNNs in reaching unprecedented energy efficiency and memory footprint with little to no compromise in accuracy.

\begin{figure}[t]
\centering
\includegraphics[width=\linewidth]{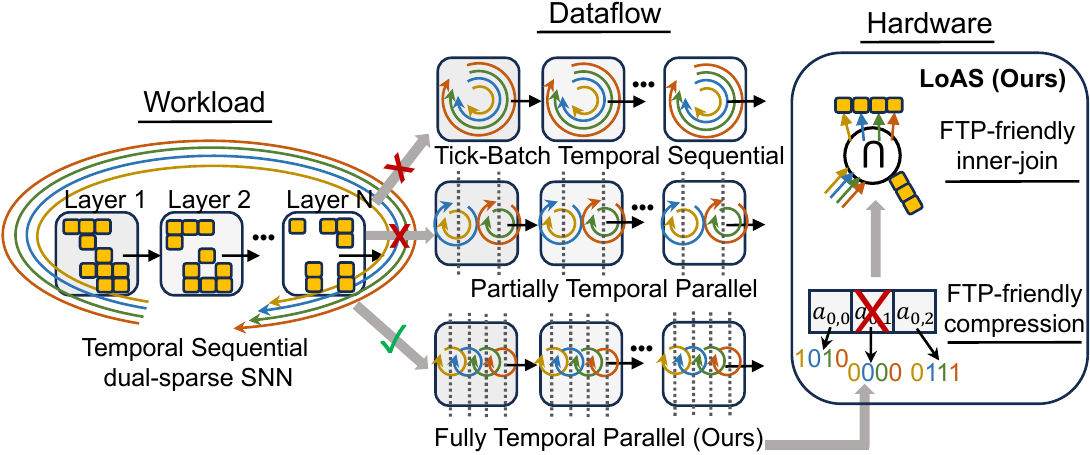}
\vspace{-3mm}
  \caption{An illustrative example of FTP dataflow and LoAS. FTP dataflow is shown along with the prior dataflow design for SNNs. Temporal sequential tick-batch is from SpinalFlow~\cite{spinalflow}, and partially temporal parallel is from PTB~\cite{ptb}. Each arrow loop indicates the processing of one timestep. The vertical line indicates that the processing is in parallel.}
  \label{fig:intro}
  
\end{figure}

\minisection{Challenge.} Although dual-sparse SNNs have made strides with algorithmic advancements, the hardware is not yet catching up to make full use of such dual-sparsity.
In general, existing SNN accelerators can be categorized into two main groups. 
First, multi-core neuromorphic systems\footnote{We are not comparing with those systems due to our focus on single-core dataflow SNN accelerator designs.} employ a plethora of cores, even chips, to exploit the inherent parallelism in spiking neuron dynamics~\cite{loihi,truenorth,spinnaker,tianji}.
Though capable of capturing the massive parallelism and sparse activities across neurons, multi-core neuromorphic systems require all neurons (including weights) to be mapped on-chip. 
This undoubtedly wastes a huge amount of hardware resources on the neurons that are not involved in any computations due to the dual-sparsity~\cite{parashar2017scnn}. 
Second, dataflow-based SNN accelerators draw inspiration from dataflow-based ANN accelerators and take advantage of the rich data reuse among the array of processing elements~\cite{ptb,spinalflow,mao2024stellar}.
Nonetheless, these designs have mainly focused on processing dense SNN workloads.
Currently, there is a lack of dataflow architectures that uniquely target dual-sparsity in SNNs. 
Table.~\ref{tab:prior_work} summarizes existing dataflow SNN accelerators.

\minisection{Insight.} Though spikes and weights have varying bitwidth, in dual-sparse SNNs, \textit{their interactions follow the pattern in sparse-matrix-sparse-matrix multiplication (spMspM)}, which has been extensively studied in ANNs~\cite{sparten,sigma,ucnn,extensor,outerspace,gospa,sparch,scnn,matraptor,gamma}.
However, naively running dual-sparse SNNs on existing spMspM accelerators is inefficient.
The reason is multifaceted.
First, the timesteps in SNNs complicate the dataflow design for existing spMspM accelerators. 
spMspM operations in ANNs are triple-nested for-loops~\cite{scnn,sparseloop}. 
Different spMspM dataflows are obtained by permuting the order of loops. 
However, in SNNs, the timesteps introduce an extra level of for loop, leading to extra latency and memory traffic.
What's worse, it constrains dataflow dependency and doubles the dataflow design space, delaying the time-to-solution.
Second, the asymmetric bitwidth of spikes and weights in SNNs makes it inefficient to use conventional compression formats in ANN spMspM accelerators.
Existing ANN spMspM accelerators store sparse matrices with popular compressed formats like compressed sparse row (CSR).
These formats usually have multiple bits to record the coordinates of the non-zero values, and so does the hardware designed.
Consequentially, using multiple bits to compress single-bit spikes (valued at either 1 or 0) is extremely inefficient for dual-sparse SNNs.

\begin{table}[t]
    \centering
    \caption{Comparison of LoAS with prior SNN accelerators. S and T denote the spatial and temporal dimensions. Spatial parallelism means PE-level parallelism.}
    \vspace{-2mm}
    \begin{adjustbox}{max width=\linewidth}
    \begin{tabular}{lcccc}
        \toprule
        Accelerator & Spike & Weight & Parallel & Neuron\\
        &Sparsity&Sparsity&support& support\\
        \midrule
        \midrule
        SpinalFlow\cite{spinalflow}& \tealcheck&\redcross &S & LIF \\
        PTB\cite{ptb}& \tealcheck & \redcross&S+partial-T & LIF  \\
        Stellar\cite{mao2024stellar}& \tealcheck & \redcross& S+fully-T& FS  \\
        LoAS (ours)& \tealcheck & \tealcheck & S+fully-T& LIF \\
        \bottomrule
    \end{tabular}
    \end{adjustbox}
    \label{tab:prior_work}
    \vspace{-5mm}
\end{table}

\minisection{Proposal.} To solve these problems and unleash the potential of dual-sparse SNNs in the presence of spMspM, we propose fully temporal-parallel (FTP) dataflow, illustrated in Figure~\ref{fig:intro}. FTP dataflow parallelizes all timesteps to avoid complicated dataflow dependency for minimized latency and memory traffic.
To maximize the efficiency of FTP dataflow on memory and computation, we design FTP-friendly spike compression and inner-joint mechanism.
The proposed compression packs spike along timesteps and can access the relevant memory space in a contiguous manner.
The proposed inner-join nearly halves the cost of cumbersome prefix-sum circuits with almost no throughput penalty compared to prior inner-join designs.
To validate FTP dataflow, we design \name, a \underline{\textbf{Lo}}w-latency Inference \underline{\textbf{A}}ccelerator for Dual-Sparse \underline{\textbf{S}}piking Neural Networks.
Our contributions are listed below:
\begin{enumerate}
\item We observe that SNNs with rich dual-sparsity from both input spikes and weight connections are sub-optimal on existing hardware. SNN hardware usually does not support sparse weights, while ANN spMspM hardware fails to efficiently process timesteps in SNNs with low latency and memory traffic.

\item To improve the efficiency of processing timesteps, we propose a fully temporal-parallel~(FTP) dataflow. 
FTP avoids extra memory traffic across timesteps and minimizes the latency penalty in processing timesteps sequentially.

\item To make the most of FTP, we propose FTP-friendly spike compression for efficient yet contiguous memory access and an FTP-friendly inner-join mechanism for low-cost computation with almost no latency penalty.

\item We build \name, a novel architecture that exemplifies the FTP dataflow.
With both FTP-friendly compression and inner-join, \name is able to achieve high speedup and energy efficiency against other sequential-running spMspM baselines.
\end{enumerate}

The remainder of the text is organized as follows. 
Section~\ref{sec:bg} reviews the background and justifies the motivation.
Section~\ref{sec:dataflow} and~\ref{sec:LOAS} articulates our proposed FTP dataflow and \name architecture. Next, Section~\ref{sec:eval_method} and~\ref{sec:eval_result} evaluate our design. Finally, Section~\ref{sec:discussion} and~\ref{sec:conclusion} discuss and conclude this work.

\section{Background and Motivation}
\label{sec:bg}

\begin{figure}[t]
\centering
\includegraphics[width=0.8\linewidth]{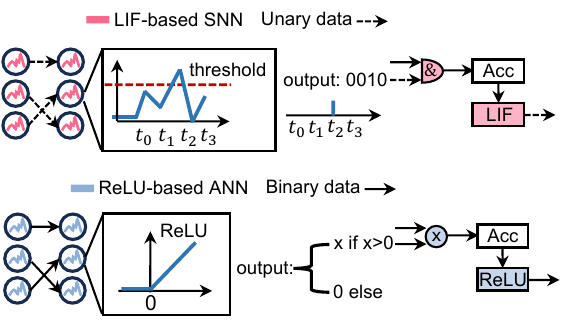}
\vspace{-3mm}
  \caption{Difference between the LIF-based SNN neuron and ReLU-based ANN neuron. We compare the behavior of LIF-based and ReLU-based neurons and their hardware implementations.}
  \label{fig:bg:snn_ann}
  \vspace{-3mm}
\end{figure}

\subsection{Preliminary of SNNs }
\label{sec:snn_preliminary}
\subsubsection{Leaky-Integrate-and-Fire Neuron}

The Leaky-Integrate-and-Fire (LIF) neuron is a classical neuron model~\cite{lif} and widely adopted by prior SNN works~\cite{tdbn,bntt,kim2022exploring,tssl}, thanks to its bio-plausibility and high accuracy. 
In this work, we focus on accelerating the workloads of dual-sparse SNNs that use LIF neurons.

During inference, each layer has an input spike tensor 
%\yk{matrix means 2-D tensor. Here, we use a 3-D tensor. The correct term is ``input spike tensor"} 
$A \in \mathbb{U}^{M\times K\times T}$ where $\mathbb{U} \in \{0,1\}$ and a weight matrix defined as $B \in \mathbb{Z}^{K \times N}$.
%\yk{{Z} means integer, {R} means real value}.
Here $T$ is the number of total timesteps; $M$, $N$, and $K$ are the spatial dimensions of the input and weight matrix. The behavior of an SNN layer can be described below:
%Assume that the matrix $A$ and $B$ are both sparse.

\minisection{Step 1: Sparse Matrix Multiplication} Sparse matrix multiplication across all timesteps is performed to obtain the full output matrix $O\in \mathbb{Z}^{M \times N\times T}$, which will be sent to LIF neurons.
\begin{equation}
    O_{m,n}[t_i]=\sum^{K}_{k=0}A_{m,k}[t_i]  B_{k,n},
\end{equation}
where the $t_i$ is the current timestep. 
With dual-sparsity, sparse matrix multiplication becomes spMspM.

\minisection{Step 2: LIF firing} LIF neurons take the snapshot of $O$ at timestep $t_i$ and generate a snapshot of the output spike tensor $C\in \mathbb{U}^{M \times N \times T}$ for current timestep $t_i$:
\begin{equation}
    C_{m,n}[t_i] = \left\{
                    \begin{array}{lccl}
                    1 & & & {X_{m,n}[t_i] > v_{th}} \\
                    0 & & & {\textnormal{else}},
                    \end{array} \right.
\end{equation}
where
\begin{equation}
    X_{m,n}[t_i]= O_{m,n}[t_i] + U_{m,n}[t_{i-1}].\notag\\
\end{equation}
Here, $U[t_{i-1}]$ is the membrane potential that carries over the temporal information from previous timestep $t_{i-1}$, and $v_{th}$ is the firing threshold, a pre-defined scalar value.

\minisection{Step 3: Membrane Potential Update} After the output spikes are generated, we update the membrane potential that will carry residual information to the next timestep according to the equation below.\footnote{We focus on the hard reset (membrane potential is reset to zero if there is an output spike of one) in this work. Though there exist other reset schemes, sticking with one of them will not lose generality in the hardware design.}
\begin{equation}
    U_{m,n}[t_{i}]= \tau X_{m,n}[t_{i}] (1 - C_{m,n}[t_{i}]),\\
\end{equation}
where $\tau \in (0,1)$ is the leaky factor.  From the above equations, we observe that to generate the output spike matrix $C$ for timestep $t_i$, we need to know the information from the previous timestep $U[t_{i-1}]$. 
This brings temporal dependency between output spike matrices across timesteps. The behavior of a LIF neuron can be found in Figure~\ref{fig:bg:snn_ann}.

\subsubsection{Spike Encoding and SNN Training}
One key step in leveraging SNNs in conventional machine learning tasks is encoding the input source data (e.g., image pixels or text embeddings) into spike trains across multiple timesteps. The input spike trains are then sequentially sent to the SNN for processing.
Recent SNN works adopt direct encoding (a special case of rate encoding) to achieve high accuracy on conventional computer vision tasks in very few timesteps ($\le4$)~\cite{direcrt_encoding,tdbn,kim2022exploring,yin2023workload, kim2022rate}. In direct encoding, the source data, instead of being directly converted into spike trains, first goes through one ANN layer. The output from the ANN layer is then converted into spike trains. We will focus on accelerating direct-coded dual-sparse SNNs in this work. The SNNs are trained using backpropagation-through-time (BPTT)~\cite{werbos1990backpropagation} with surrogate gradient~\cite{neftci2019surrogate} to achieve very close performance to ANNs on many complex tasks~~\cite{direcrt_encoding,tdbn}.

\subsection{Distinctive Features and Challenge of SNNs}
\label{sec:bg:snn_con}
Several distinctive features make SNNs favorable for low-power edge deployment, but they also come with challenges.

\minisection{Feature 1: Unary Activation} One of the most distinctive features of SNNs is their unary spike activation. More specifically, the SNNs leverage single-bit non-weighted activation to propagate information through layers. 
The primary benefit of the unary activation is the simplified low-power arithmetic units that they require. As shown in Figure~\ref{fig:bg:snn_ann}, compared to the multiply-accumulate (MAC) of ANNs, SNN only requires simple bitwise-AND and accumulate (AC) operations during inference time.\footnote{There exist other implementations using multiplexers instead~\cite{ptb,spinalflow}. We focus on using bitwise-AND gates in this work.} Without the expensive multipliers~\cite{han2016eie}, the computations for SNNs require extremely low power and area.

\minisection{Feature 2: Sparse Spike Activity} The second feature of SNNs is their highly sparse spike-firing activity. In ANNs, upon completion, MAC results go through the ReLU unit, which filters out non-positive outputs. Different from ANNs, AC results in SNNs go through the Leaky-Integrate-and-Fire (LIF) unit, which only fires (generates an output of 1) when the input is greater than a pre-set threshold. As a result, the output sparsity in SNNs is usually much higher ($\sim90\%$)~\cite{tssl,yin2024workload,yin2024mint,tdbn} than that of ANNs ($\sim50\%$)~\cite{scnn,bg_ann_spa}. More sparse outputs apparently lead to more computation and memory saving under the context of spMspM acceleration.
%Table showing sparsity comparison.

\label{sec:bg:timestep}
\minisection{Challenge: Repeated Timesteps} Despite the aforementioned hardware-friendly features, one main challenge of deploying SNNs on hardware is their intrinsic repeated timesteps. 
A timestep is the minimum unit of time in SNNs, thus discrete.\footnote{Timestep is also called tick~\cite{spinalflow} or time-point~\cite{ptb} in other works. 
We follow the naming convention adopted by the latest SNN algorithm works.}
In one timestep, each neuron needs to complete the AC operations for all inputs, fire a spike if necessary, and update its membrane potential (will be discussed shortly). 
The SNN needs to run across multiple timesteps to capture the temporal dynamics from the input data, as shown in Figure~\ref{fig:bg:snn_ann}. 
Running multiple timesteps increases latency and fails to be energy efficient, diluting the advantage of low-power circuits unless we have a specialized architecture design~\cite{spinalflow}.

\begin{figure*}[h]
\centering
\includegraphics[width=0.9\linewidth]{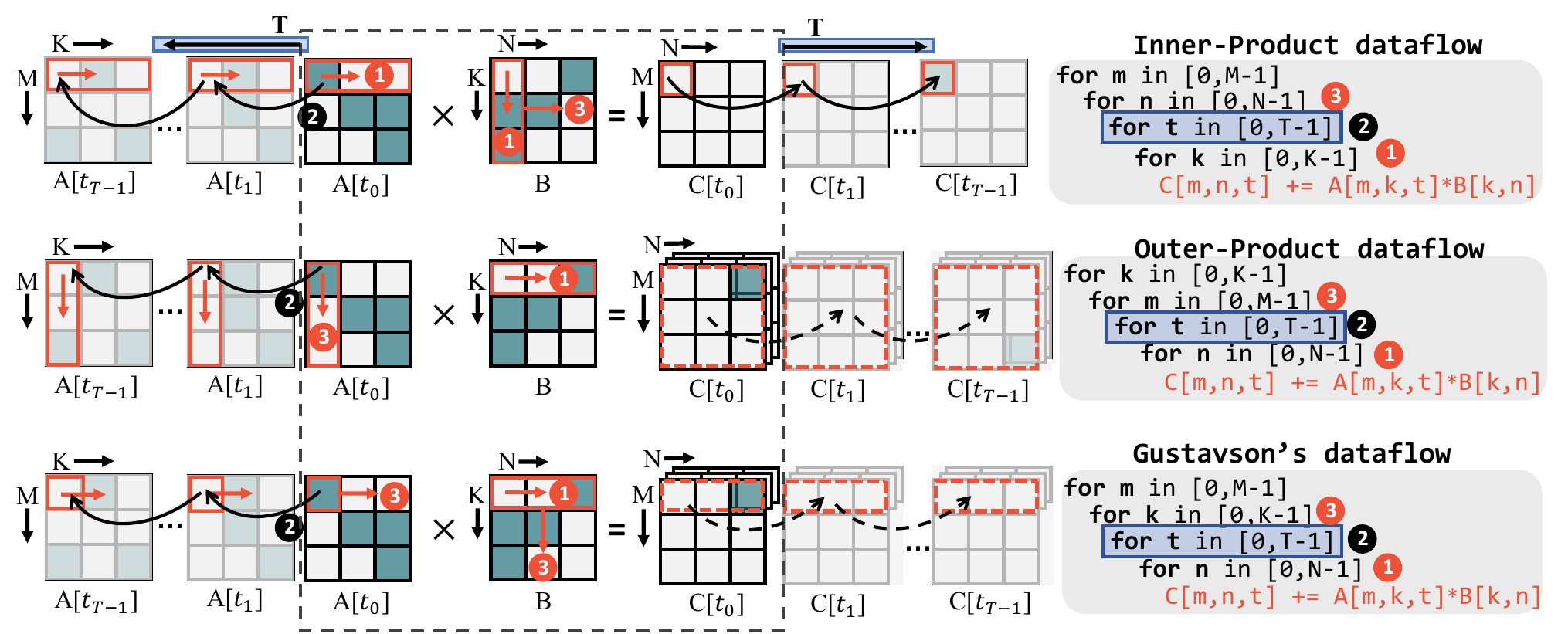}
\vspace{-3mm}
  \caption{Comparison of different spMspM dataflow for SNNs. Here, for illustration purposes, we put $C[t_i]$ as the spMspM result between $A[t_i]$ and $B$ to align with spMspM in ANNs. In SNNs, we need to go through one more LIF step (Equation~(2)) to get $C[t_i]$. The circled numbers illustrate the order of computation for the specific spMspM dataflow. Please note that we fix the position of \textbf{\texttt{t}} dimension for illustration purposes. In practice, there will be a total of 16 possible permutations of spMspM dataflow in SNNs, which we will discuss in Section~\ref{sec:dataflow}.}
  \label{fig:bg:dataflow}
  \vspace{-3mm}
\end{figure*}

\subsection{spMspM Dataflows in SNNs}
\label{sec:spm_dataflow}
% Unlike the spMspM dataflow in ANNs, which is based on a triple-nested for loop, the spMspM dataflow in SNNs needs to consider one more loop
There are various ways to map spMspM onto hardware, each with unique efficiency~\cite{spada,flexagon}. 
Three different spMspM dataflows have been proposed in existing dual-sparse ANN accelerators: Inner-product (\textbf{\texttt{IP}})~\cite{sparten,sigma,ucnn,extensor}, Outer-product (\textbf{\texttt{OP}})~\cite{outerspace,gospa,sparch,scnn}, and Gustavson’s (\textbf{\texttt{Gust}})~\cite{matraptor,gamma}.  
In Figure~\ref{fig:bg:dataflow}, we illustrate these three dataflows in SNNs for two input matrices $A$ and $B$, and an output matrix $C$. 
% Inside the box, all dataflows are identical to the ones in ANNs. 
We also formulate their abstract loop nests on the right-hand side. 
As we discussed in Section~\ref{sec:bg:timestep}, it is impossible not to consider the multiple timesteps for spMspM operations in SNNs. 

Inside the black box in Figure~\ref{fig:bg:dataflow}, the dataflow is for one timestep, thus identical to ANN dataflow.
Outside the black box, multiple input matrices $A$ (blurred) represent the input spike matrices across different timesteps, which need to be processed. 
Meanwhile, multiple output spike matrices $C$ that have temporal dependency between each other are also generated. 
Specifically, to accommodate the timesteps in SNNs, we need to consider one more loop dimension ($t$ dimension) in the original triple-nested for-loop. 
The $t$ dimension (annotated in the blue box) brings temporal dependency to each output pixel in SNNs. 
For example, to process the SNN using \textbf{\texttt{IP}} dataflow as shown in Figure~\ref{fig:bg:dataflow}, we first calculate the output cell at (0,0) position for timestep $0$ ($C$[0,0,0]), then instead of moving to the position (0,1), we move on to process the output cell at (0,0) for timestep $1$ ($C$[0,0,1]). 
Since the output cell $C$[0,0,1] is temporal dependent on the result of the output cell $C$[0,0,0], we cannot process $C$[0,0,1] before $C$[0,0,0].

\subsection{ANN spMspM Hardware for dual-sparse SNNs}
\label{sec:spmspm_snn}

We review existing ANN spMspM accelerators to understand why naively running dual-sparse SNNs on these accelerators is sub-optimal.

\begin{figure}[t]
\hspace{5mm}
\includegraphics[width=0.7\linewidth]{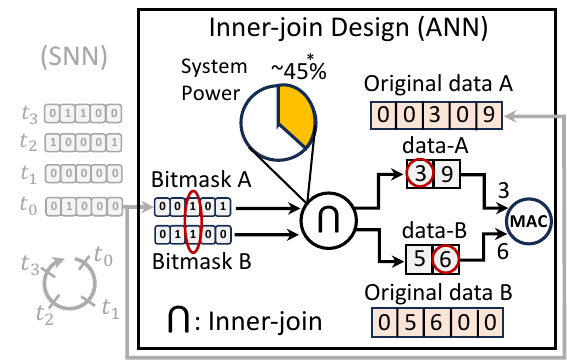}
\vspace{-3mm}
  \caption{An example of the inner-join design. The difference between the behavior of ANN and SNN is shown. *Data from SparTen~\cite{sparten}.}
  \label{fig:bg:inter}
   \vspace{-5mm}
\end{figure}

\minisection{Inner-join Design}: For the \textbf{\texttt{IP}}  dataflow, prior accelerators usually adopt the inner-join-based design~\cite{sparten,gospa}. 
In such designs, non-zero values in rows of matrix A and columns of matrix B are compressed using bitmask representation (a bit string that has 1’s for positions with non-zero values and 0’s otherwise).
An inner-join unit scans two bitmasks on the fly to determine if there's a matched position (both multiplicands are non-zero) and then sends the matched pairs to the compute units. Running dual-sparse SNNs on an inner-join-based design does not require the extra bit-masks for the input spike matrix $A$ (the unary spike train itself can be viewed as a bit-mask). However, as shown in Figure~\ref{fig:bg:inter}, the timesteps will impose multiple extra rounds of running the expensive inner-join units (e.g., occupying roughly 46\% of the system-level power~\cite{sparten}), thus incurring high energy cost. Moreover, since the spike trains are used as bit-masks, all the spikes, no matter 1 or 0, are necessary to be fetched from off-chip DRAM. This brings no memory traffic saving on the sparse spike matrix $A$.

\begin{figure}[h]
\centering
\includegraphics[width=0.7\linewidth]{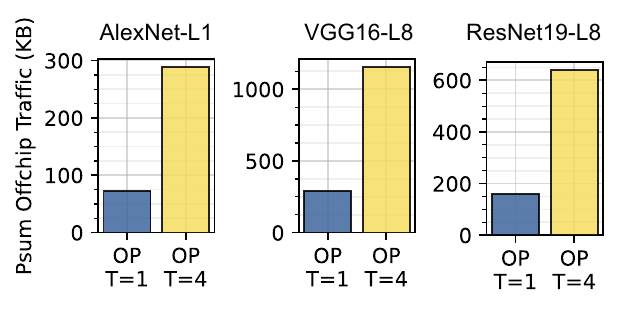}
\vspace{-5mm}
  \caption{Off-chip traffic of partial sum matrices on different SNN layers. We envision SNNs with a timestep of 1 and 4 running on GoSPA~\cite{gospa}, an OP dataflow spMspM accelerator.}
  \label{fig:bg:simple_op}
   \vspace{-3mm}
\end{figure}
\minisection{Merger-based Design}: Unlike \textbf{\texttt{IP}}  dataflow designs that exhibit full output matrix ($C$) reuse, \textbf{\texttt{OP}} and \textbf{\texttt{Gust}} dataflow designs focus on the reuse of input matrix $A$ and $B$. 
In \textbf{\texttt{OP}}, each column of $A$ and each row of $B$ will only be transversed once, leading to efficient input data reuse. 
However, one partial sum is generated at a time and merged later. While these two dataflows have better data reuse on the input matrix, the partial sum matrices (rows) potentially bring more off-chip data traffic. To amortize the large memory traffic of partial sums, some designs implement large and costly mergers (e.g., 38$\times$ more area than multipliers~\cite{sparch}) to merge as many as partial sum matrices (rows) before sending them back to the off-chip DRAM.
Due to the extra $t$ dimension, running dual-sparse SNNs on a merger-based design either requires a more complex merger that is capable of digesting the extra partial sum traffic or incurs more off-chip memory traffic. As shown in Figure~\ref{fig:bg:simple_op}, for a timestep of four, on average, $4\times$ more partial sum traffic will be induced compared to a single timestep.

\subsection{Dataflow Architecture for SNNs}
\label{sec:bg:prior_snn}

\minisection{SpinalFlow: Temporal Sequential Design.} 
SpinalFlow~\cite{spinalflow} is the first SNN-tailored accelerator for extracting the efficiency from the single-bit activation and the extremely sparse spike activity. 
The authors identified the challenge of sequentially processing the entire SNN network through timesteps. 
To overcome the challenge, SpinalFlow proceeds all timesteps for one layer and then proceeds to the next layer, as shown in Figure~\ref{fig:intro}. 
SpinalFlow dispatches LIF neurons across different processing elements (PEs) and parallelizes the computation.
Within each layer, the timesteps are processed sequentially, as shown in Figure~\ref{fig:intro}. 
Spinalflow is optimized exclusively for the temporal-coded SNNs that potentially lag in terms of accuracy performance compared to rate-coded SNNs~\cite{ptb}.
In this work, we focus on accelerating spMspM for general rate-coded SNNs that yield competitive accuracy as ANNs in various tasks.

\begin{figure}[t]
\centering
\includegraphics[width=0.85\linewidth]{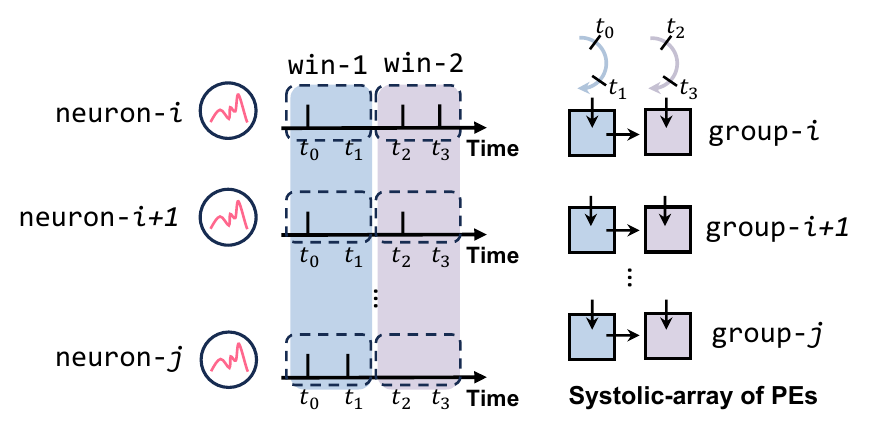}
\vspace{-3mm}
  \caption{Example of PTB's partially temporal parallel design. Each column of the PE array processes a time-window that consists of multiple timesteps. Different time-windows run in parallel, but the timesteps inside the window are still processed sequentially.}
  \label{fig:rw:ptb}
   \vspace{-5mm}
\end{figure}

\minisection{PTB: Partially Temporal Parallel.} While SpinalFlow's design is tailored to the temporal-coded SNNs, PTB~\cite{ptb} proposes a general architecture design for the rate-coded SNN. 
By leveraging the high data-reuse pattern across different PEs in the systolic array architecture~\cite{kung1982systolic}, PTB breaks the processing of all timesteps into multiple time-windows (each consists of several contiguous timesteps) and run these time-windows in parallel, as shown in Figure~\ref{fig:intro}.
PTB parallelly maps multiple time-windows across different columns of the systolic array. The computation of different LIF neurons is also parallelized across the rows of the systolic array. We illustrate this hardware mapping strategy in Figure~\ref{fig:rw:ptb} with details. Though PTB tries to parallelize the processing of timesteps, the parallelization is on the granularity of the time-window. Inside each time-window (column of PEs), the timesteps are still processed sequentially. Consequently, we categorize PTB as a partially temporal parallel design. 
One unique aspect of \name from PTB is that \name places the temporal dimension in the inner-most loop, enabling all optimizations.

Prior SNN accelerators with LIF neurons process timesteps in a sequential or partially parallel manner. In this way, as we discussed in (Section \ref{sec:spm_dataflow} \& \ref{sec:spmspm_snn}), it is very challenging for those existing SNN designs to have good performance on spMspM SNN acceleration.
Thus, we need a spMspM-friendly strategy to process timesteps. 

\minisection{Stellar: Fully Temporal Parallel but with non-LIF neurons.} Stellar~\cite{mao2024stellar} is another systolic array SNN accelerator which attempts to process timesteps in a fully parallel manner. Nonetheless, Stellar focuses on optimizing for the Few Spikes (FS) neuron~\cite{fs}, as shown in Table~\ref{tab:prior_work}. FS neurons behave differently from LIF neurons by detaching the spike accumulating and firing stages. Therefore, FS neurons naturally do not have temporal dependency among the input data at the spike accumulation stage. This makes fully parallel temporal processing straightforward in Stellar. On the contrary, as discussed in Section~\ref{sec:snn_preliminary}, temporal dependency naturally exists in the input data for the LIF neuron, which makes its design space different from the one in Stellar for fully temporal parallel processing. Unlike the widely adopted LIF neurons, supporting FS neurons also requires non-trivial algorithm-hardware codesign, which is out of the scope of this work.

\begin{algorithm}[t]
\caption{Fully Temporal-Parallel dataflow (\textbf{\texttt{FTP}})}\label{alg:tp_dataflow}
       \textbf{Input}: \\
       Input spike matrix $A \in \mathbb{U}^{M\times K \times T}$($\mathbb{U} \in \{0,1\}$)\\
       Weight matrix $B \in \mathbb{Z}^{K\times N}$\\
      \textbf{Output}: \\
      Output spike matrix $C \in \mathbb{U}^{M\times N \times T}$
      \begin{algorithmic}[1]
        \For{$m \in M$}
        \For{$n \in N$}
        \For{$k \in K$}\\
        \indent \indent \indent \textbf{parallel-for} $t \in T$ \textbf{do} \Comment{Spatially unrolled}
        \State\hspace{3mm}$O[m,n,t]\text{ += } A[m,k,t]\times B[k,n]$
        \EndFor
        
        \indent \indent \textbf{parallel-for} $t \in T$ \textbf{do}
        \Comment{Spatially unrolled}
        \State$\indent C[m,n,t]\text{ = } LIF(O[m,n,t])$
        \EndFor
        \EndFor
      \end{algorithmic}
\end{algorithm}
\section{Fully Temporal Parallel Dataflow}
\label{sec:dataflow}
We propose a \textit{fully temporal-parallel dataflow}~(\textbf{\texttt{FTP}}) that targets reducing the negative effects of repeatedly processing the timesteps on spMspM accelerators (Section~\ref{sec:spmspm_snn}).
The proposed \textbf{\texttt{FTP}} is formulated in Algorithm~\ref{alg:tp_dataflow}.

An SNN-friendly spMspM dataflow should satisfy three goals: (1) avoid as much data refetch as possible across the timesteps; (2) generate as few partial sums as possible on the temporal dimension (timesteps); (3) reduce the latency as much as possible on the temporal dimension to reduce the extra cost of sparsity handling units. 

Our first observation is that for all three spMspM dataflows (Section~\ref{sec:spm_dataflow}), unless placing the temporal dimension ($t$-dim) at the innermost loop, it will bring at least $T$ times more data refetch to the dimensions below, compared to the original dataflow. For example, in \textbf{\texttt{OP}}, if $t$-dim is placed between $m$ and $n$, $T$ times more access to $B$'s rows is required. If $t$-dim is placed between $k$ and $m$, $T$ times more access to $A$'s columns and $B$'s rows is required. Depending on the on-chip buffer capacity, repeated memory access might lead to more expensive access to the off-chip memory, which opposes goal~(1).

Our second observation is that both \textbf{\texttt{OP}} and \textbf{\texttt{Gust}} dataflow are not suitable for dual-sparse SNNs since they oppose goal~(2). In \textbf{\texttt{OP}} dataflow, we observe that no matter where we insert the $t$ dimension into the original triple-nested loop, we always produce $T$ times more partial sum matrices compared to the original \textbf{\texttt{OP}} dataflow. The partial sums need to be stored in an on-chip cache till all partial sums along both spatial ($k$) and temporal dimensions ($t$-dim) are accumulated. This will add extra memory overhead in \textbf{\texttt{OP}}. 
The same problem also exists for \textbf{\texttt{Gust}} dataflow. The $t$-dim will either generate $T$ times more partial sum rows or have $T$ times more access to both $k$ and $n$ dimensions. The last observation is that regardless of the position of $t$-dim, as long as we process it sequentially, it always incurs $T$ times more processing latency, which opposes goal~(3).

Our solution is straightforward but effective. We first choose to position the $t$-dim at the innermost of the \textbf{\texttt{IP}} dataflow, as given in Algorithm~\ref{alg:tp_dataflow}.
This design choice has several advantages. 
Firstly, putting the $t$-dim at the innermost loop ensures that no extra data movement will be incurred (goal~(1)). 
Secondly, since \textbf{\texttt{IP}} dataflow has efficient output reuse, no extra partial sums will be generated on the $t$-dim (goal (2)). 
Lastly, we fully parallelize the $t$-dim and eliminate the latency brought by sequentially processing timesteps.
This is equivalent to transforming the \textit{for-loop} of $t$ into a \textit{parallel-for} loop~\cite{sparseloop}. 
This \textit{parallel-for} loop parallelizes the operation across different spatial instances, requiring minimum hardware overheads due to only cheap accumulators being duplicated, and timesteps of direct-coded SNNs are small (Section~\ref{sec:snn_preliminary}). We later show in the ablation studies that \textbf{\texttt{FTP}} scales well with the increasing timesteps.
\begin{figure}[h]
 \centering
 \vspace{-4mm}
\includegraphics[width=\linewidth]{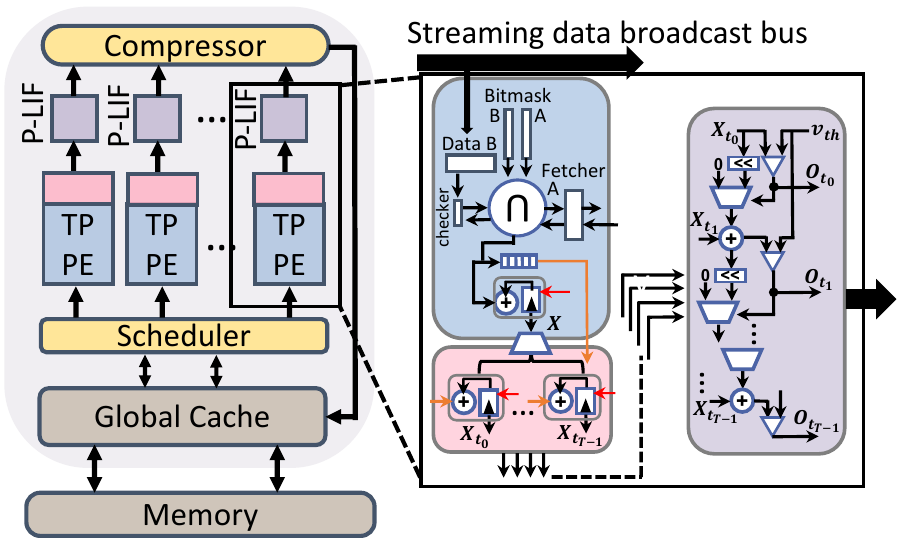}
 \vspace{-4mm}
\caption{Architecture of \name and the microarchitecture of the TPPE. Red arrows are the enable signals that skip the computation on 0 spikes~\cite{sata}.}
\label{loas:micro-arch}
 \vspace{-5mm}
\end{figure}
\section{\name}
\label{sec:LOAS}

An overview of {\name} is shown in Figure~\ref{loas:micro-arch}. {\name} consists of multiple {temporal parallel processing elements (TPPEs)} and {parallel Leaky-Integrate-Fire units (P-LIFs)} that are tailored to run the FTP dataflow; a {scheduler} that distributes workloads across TPPEs; and a {compressor} that compresses the output spikes from P-LIFs and writes them back to the on-chip memory. An on-chip SRAM is equipped to capture data reuse.

\subsection{Spikes Compression}
\label{sec:mem_compress}

We first discuss how sparse input spikes (matrix $A$) across timesteps are compressed in \name. Efficiently compressing matrix $A$ in SNNs necessitates solving two challenges:

\minisection{How to maximize the compression ratio of 1-bit spikes}?
Assume that the input spike matrix $A$ has a size of 128$\times$128 for each timestep. 
Then for either CSR or CSC, we need to use two 7-bit coordinates to compress each 1-bit non-zero spike.\footnote{For 128 columns, we need $\log_2(128)=7$ bits for coordinates. We neglect the offsets in the discussion, which will further increase the number of bits used for coordinates.}
Furthermore, SNNs naturally run for multiple timesteps, which means that for the same coordinate, different spike values may occur at different timesteps (e.g., $0$ for T=1\&3, and $1$ for T=2\&4). To faithfully capture all the non-zero spikes, we need separate coordinate values for each timestep.

\minisection{How to maintain contiguous memory access of non-zero spikes across timesteps}?
The \textbf{\texttt{FTP}} dataflow we proposed in Section~\ref{sec:dataflow} requires spatial unrolling of the input spike matrix $A$ across all timesteps beneath the $k$ dimension. 
Consequently, a dis-contiguous memory layout of $A$ along the $t$ dimension will cause fragmented memory access at all levels of memory hierarchies, leading to higher data movement costs.
\begin{figure}[t]
\centering
\includegraphics[width=0.75\linewidth]{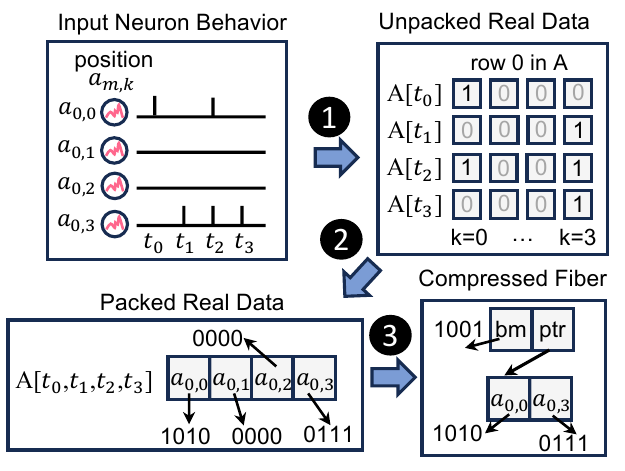}
\vspace{-3mm}
  \caption{Example showing how input spikes are compressed in \name. bm stands for the bitmask, and ptr stands for the pointer.}
  \label{fig:arch:mem_compress}
   \vspace{-5mm}
\end{figure}

To better illustrate these two points, we provide an example in Figure~\ref{fig:arch:mem_compress}. Envisioning that the input spikes sent to the system have the pre-synaptic neuron $a_{0,0}$ (first element of row-0 in matrix $A$) firing a spike at $t_0$ and $t_2$. As shown in step \circled{1}, to represent this pre-synaptic neuron behavior, a single-bit 1 needs to be stored at row-0, column-0 of matrix $A$ for both timestep 0 and 2 into the memory, shown in the box of 'unpacked real data.' Then, for each non-zero spike in row-0 of matrix A for each timestep, if we need to use a coordinate value (e.g., 4-bit for CSR) to record its position. We then need $2\times4=8$ bits to compress 2 bits (2 spikes). The compression efficiency in this case is only $25\%$. Furthermore, memory access to spikes across different timesteps is discontinuous (sequentially access different rows of $A$).
We propose the following spikes compression format for \name to solve these two challenges. 
In our method, as shown in step \circled{2}, we pack all the spikes (both 0 and 1) across all timesteps into one continuous data block in the system for each pre-synaptic neuron.
In the example of Figure~\ref{fig:arch:mem_compress}, we store a 4-bit value 1010 at the first position of row-0 of matrix $A$ for $a_{0,0}$ and 0111 at the fourth position for $a_{0,3}$. 
Since neurons $a_{0,1}$ and $a_{0,2}$ do not spike at any timestep, their packed value would be 0000 (shown in the box of 'packed real data'). We define these neurons as silent neurons.\footnote{We follow the same terminology used in~\cite{ptb}.}
With this strategy, only the non-silent neurons will be treated as non-zero values and stored in the memory for matrix $A$, as shown in step \circled{3}. In our example, we end up using 4 bits to compress 5 bits. The compression efficiency in this case is $125\%$.

To accommodate our \textbf{\texttt{FTP}} dataflow, we compress the input spike matrix $A$ in a row-wise manner and use the bitmask format~\cite{sigma,sparten,gospa} to represent the coordinates of the non-zero values. 
The bitmask format uses a 1-bit coordinate value for each position in the row. In our example, the bitmask is 1001 since the first and the fourth elements in the row are non-zero. The second and third elements are silent neurons, so we do not store them in the memory (represented by a 0 in bitmask). Following the bitmask, a pointer is stored to provide the starting location of the non-zero values of the row. We call this compressed row: a fiber~\cite {gamma,flexagon}.

The key to our compression method is the ratio of silent neurons in the SNN. Fortunately, empirical studies have shown that SNNs have a significant fraction of silent neurons ($60\%\sim 70\%$, as shown in Table~\ref{tab:workload}).
We further use a similar bitmask-based technique to compress weights in a column-wise manner. Each compressed weight column is also called a {fiber}.

\subsection{Temporal Parallel Processing Elements}

The fundamental building blocks of \name's compute engine are Temporal Parallel Processing Elements (TPPEs) and Parallel Leaky-Integrate-Fire units (P-LIFs), which we describe next. Figure~\ref{loas:micro-arch} also details the design of TPPE. Each TPPE produces the full sum for one output neuron across all timesteps (Line 5 in Algorithm.~\ref{alg:tp_dataflow}). Before the computation starts, the bitmask (bm-B) of a fiber from weight matrix $B$ (fiber-B) and its non-zero data are read from SRAM and broadcasted into the small bitmask buffers (128 bits in our design) inside each TPPE.
The bitmask (bm-A) of fiber from input spike matrix $A$ (fiber-A) is also fetched and sent to the TPPEs. Each TPPE will hold the bitmask for a distinct fiber along the row of $A$.
After the data are loaded, an \textit{inner-join} operation~\cite{sparten,gospa,extensor} is performed between the two bitmasks.
Depending upon the inner-join result, the matched non-zero data of fiber-A will be fetched from the global cache and sent to the \textit{pseudo-accumulator} (soon be discussed) to perform the accumulation (AC) operation.
After the TPPE completes the full computation of one output neuron, it will send the result to the P-LIF unit to generate output spikes for all timesteps in one shot.

\begin{figure}[t]
\centering
\includegraphics[width=0.9\linewidth]{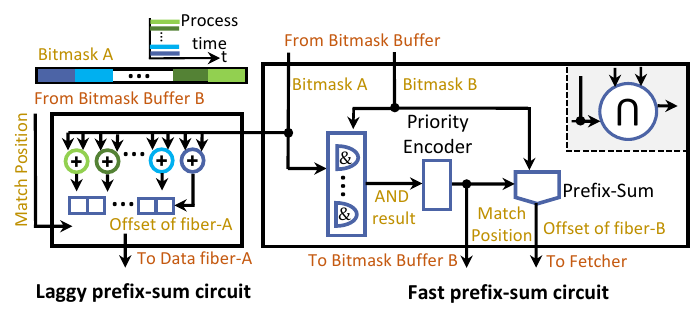}
\vspace{-5mm}
  \caption{Illustration of the proposed FTP-friendly inner join unit.}
  \label{fig:arch:inner}
   \vspace{-5mm}
\end{figure}

\subsection{Inner-join Unit}

The inner-join operation has been extensively studied by prior works~\cite{sparten,gospa,extensor} for spMspM acceleration in ANNs. The inner-join mechanism with prefix-sum circuit has been efficiently implemented with the bitmask representation~\cite{sparten}. 
In~\cite{sparten}, a logical-AND operation is first applied to two bitmasks to get the \textit{AND-result}, which represents the location where both data are nonzero. 
The \textit{AND-result} is then sent to a priority encoder to convert the \textit{matched positions} into integer values. 
The \textit{matched positios} are sent to two separate prefix-sum circuits to get the number of 1s in front of the \textit{matched position} for each bitmask. This gets the offsets for each non-zero data in the memory.

During the above process, the use of two fast prefix-sum circuits is an expensive operation (taking more than 45\% power and area in \cite{sparten}).\footnote{In~\cite{sparten}, the design of the prefix-sum circuit is not described. We assume it to be a tree-like prefix-sum circuit with O($\log(n)$) complexity that can run in one clock cycle. $n$ is the size of input and output for the prefix-sum circuit, which is set to 128 in both~\cite{sparten} and our work.}
To reduce the overhead brought by the prefix-sum circuits, we propose an FTP-friendly inner-join unit that is detailed in Figure~\ref{fig:arch:inner}.

We first observe that in ANNs, the MAC operation requires both inputs to be explicitly known at computation time. Therefore, we need two fast prefix-sum circuits to match the processing speed between two inputs. 
However, this is not the case with SNNs. In SNNs, we only have two cases for the input ($1$ or $0$), meaning we either accumulate or discard the weight.
This provides the opportunity to have an imbalanced processing speed for two inputs at the prefix-sum stage.

In our design, instead of using two fast prefix-sum circuits as in ANNs, we have one fast and one laggy prefix-sum circuit, as shown in Figure~\ref{fig:arch:inner}. 
Recall that our compression method only fetches the non-silent neurons (that fire at least once across timesteps) from DRAM for $A$. Thus, as soon as we find a matched position in \textit{AND-result}, we are confident that the corresponding non-zero value in fiber-B will be accumulated at least once (at least one timestep). Therefore, we can begin accumulating the non-zero value in fiber-B without knowing the exact spike information from fiber-A. In this way, we can ensure the throughput of consuming fiber-B is always high regardless of the processing speed of fiber-A.

In our efficient inner-join unit, each time the fast prefix-sum circuit generates an offset, the corresponding non-zero value of fiber-B will be directly sent to a \textit{pseudo-accumulator} for accumulation.
This mechanism opportunistically presumes the matched non-zero value of fiber-A is all 1s (pre-synaptic neuron fires at all timesteps) to fully leverage the throughput of the fast prefix-sum circuit. Since the non-zero value in fiber-A is not always all 1s, we need a mechanism to ensure that the accumulation results are correct. Instead of using the expensive fast prefix-sum circuit to access and check the matched non-zero value in fiber-A, we use a much simpler circuit to generate the offset of fiber-A.
We defined the simpler prefix-sum circuit as the \textit{laggy prefix-sum circuit}, illustrated on the left of Figure~\ref{fig:arch:inner}. 
We use a group of adders to sequentially add up the prefix-sum results and store them inside a small buffer. These adders run in parallel, and hence, the latency of generating all the offsets is equal to len(bm-A)/\# of adders.

\begin{figure}[h]
\centering
\vspace{-2mm}
\includegraphics[width=\linewidth]{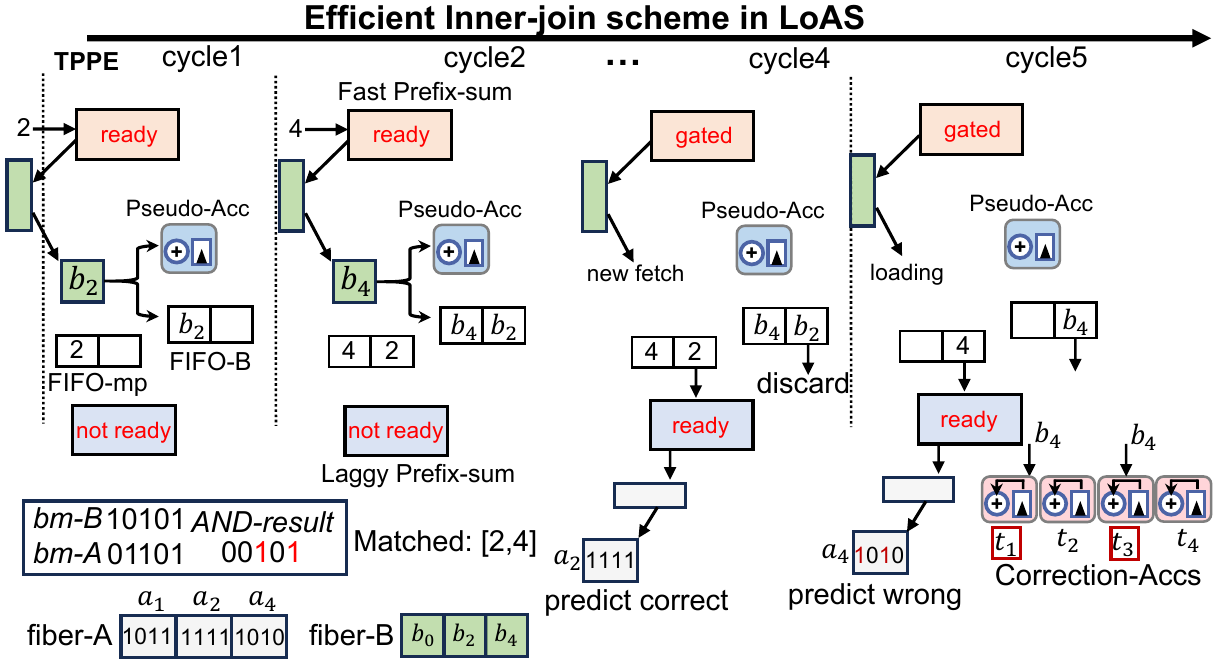}
\vspace{-5mm}
  \caption{A walk-through example of the proposed FTP-friendly inner-join unit. This example assumes the laggy prefix-sum circuit will be ready after 2 cycles. FIFO-mp is the FIFO to buffer the matched position. FIFO-B is the FIFO to buffer the matched non-zero value of B. Acc stands for accumulator.}
  \label{fig:arch:demo}
   % \vspace{-3mm}
\end{figure}

We provide a simple walk-through example in Figure\ref{fig:arch:demo}. We first run the fast prefix-sum circuit; in every cycle, we accumulate the matched non-zero value of fiber B and buffer it together with the matched position in small FIFOs. When the laggy prefix-sum circuit finishes running, a ready signal is sent out. 
We then check the non-zero value in fiber-A according to the buffered position from FIFO-mp. If the matched value is all 1s, we simply discard the current value in FIFO-B. Otherwise, we need to send the buffered non-zero values of fiber-B from the FIFO-B to the correction accumulators. As illustrated in Figure~\ref{fig:arch:demo}, at cycle 4, we check $a_2$ and find its value is 1111. Thus, we simply discard $b_2$. At cycle 5, we check $a_4$ and find its value is 1010. Thus, we send $b_4$ to the correction accumulator for $t_1$ and $t_3$. This example shows the motivation and benefits of using a combination of fast and laggy prefix sums. By having a fast prefix sum, we can consume B at the earliest possible by first accumulating it into the pseudo-accumulator. While waiting for the laggy prefix sum to correct the accumulation results, we can proceed to fetch the next fiber-B's data into the buffer. This way, the latency of fetching fiber B can be overlapped with the laggy prefix sum and correction to improve the overall throughput. At the same time, replacing one fast prefix sum with a laggy one saves the overall power and area of our TPPE.

\subsection{Other Units}
\label{sec:other}
After the computation of the \textit{pseudo-accumulator} completes, its accumulation results are duplicated and sent to each correction accumulator. The correction value inside each accumulator will be subtracted from the pseudo accumulation results for each timestep. Finally, we send the corrected results to the P-LIF units to generate the output spikes. As shown inside the purple box in Figure~\ref{loas:micro-arch}, we spatially unroll the LIF operations so that the output spikes for all timesteps will be generated at once.

\name uses a unified global buffer for holding compressed fiber-A and fiber-B with their bitmask representations. 
We adopt a FiberCache design~\cite{gamma}. A unified shared cache exhibits better utilization compared to separate ones. Each line in the global cache consists of two parts. The first part is the bitmask representation of a fiber, followed by a pointer. The second part is the non-zero values of that fiber. If the line manages to hold all the non-zero values, the pointer will be a NULL pointer. Otherwise, it will point to the location of the line where the rest of the data are held. Each PE will take responsibility for generating one output neuron. Therefore, we use a highly banked global cache to ensure multiple PEs can access their data concurrently. Inside each bank, we fetch as many chunks as possible for one fiber in matrix $A$ and hold them as long as possible to maximally have the data reuse of $A$. This can be achieved by adopting a replacement policy for the global cache as in ~\cite{gamma,spada}. Only one compressed row fiber of matrix $B$ is fetched into the global cache and broadcasted to all TPPEs. We follow a compression unit as~\cite{sparten}, where an inverted prefix-sum circuit is used to compress the output spikes and generate their bitmask representations. Similar to the observation in~\cite{sparten}, this compression step does need to be performed fast. Therefore, we equip an inverted \textit{laggy prefix-sum} circuit to perform the compression. The scheduler will be responsible for casting the data to each TPPE through a simple swizzle-switch-based crossbar~\cite{sewell2012swizzle}.

\section{Experimental Methodology}
\label{sec:eval_method}

\begin{figure}[h]
\centering
\includegraphics[width=0.6\linewidth]{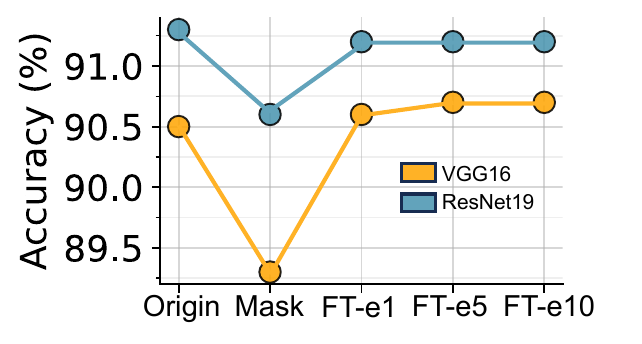}
\vspace{-5mm}
  \caption{Accuracy trends of the fine-tuned preprocessing. Mask means masking out all presynaptic neurons that fire only once during the inference. FT-ex means fine-tuning for x epochs.}
  \label{fig:exp:fine-tune}
   \vspace{-3mm}
\end{figure}

\minisection{Software Configuration: }
For the dual-sparse SNNs, we train and compress the AlexNet~\cite{alexnet}, VGG16~\cite{vgg}, and ResNet19~\cite{resnet}. 
We use the open-source toolchains for lottery-ticket-hypothesis (LTH)-based SNN pruning~\cite{lth, frankle2018lottery}. We set the default timesteps $T$ to 4 across all experiments. We use 15 rounds of LTH searching, and all SNNs are trained towards convergence with similar accuracy as state-of-the-art dense baselines~\cite{lth}. 
We further select representative layers from each network to provide single-layer insights. The summary of the workloads is in Table~\ref{tab:workload}.
\begin{table}[h]
    \centering
    \caption{SNN workloads. NL = \# of layers. T = Timesteps. AvSp\{A, B\} = Average sparsity of the matrices\{A, B\} in(\%). AvSpA-origin is the original spike sparsity across timesteps, AvSpA-packed is the density of silent neurons, and AvSpA-packed+FT is the density after fine-tuned preprocessing. M/N/K denotes matrix shape.}
    \vspace{-2mm}
    \begin{adjustbox}{max width=\linewidth}
    \begin{tabular}{lccccc}
        \toprule
        SNN & NL & T & AvSpA & AvSpA & AvSpB\\
        &&&origin&packed(+\textbf{FT})&\\
        \midrule
        \midrule
        AlexNet(A)& 7 & 4 & 81.2&71.3(\textbf{76.7}) & 98.2 \\
        VGG16(V)& 14 & 4 & 82.3&74.1(\textbf{79.6}) & 98.2  \\
        ResNet19(R)& 19 & 4 & 68.6& 59.6(\textbf{66.1})& 96.8  \\
        % \
        \midrule
        \midrule
        Layer & \multicolumn{2}{c}{T,M,N,K} & & & \\
        \midrule
        A-L4 & \multicolumn{2}{c}{4,64,256,3456} & 75.8&63.2(\textbf{69.7}) & 98.9 \\
        V-L8 & \multicolumn{2}{c}{4,16,512,2304} & 88.1&76.5(\textbf{86.8}) & 96.8  \\
        R-L19 & \multicolumn{2}{c}{4,16,512,2304} & 57.9&51.4(\textbf{55.7}) & 99.1\\
       {T-HFF} & \multicolumn{2}{c}{4,784,3072,3072} & -&- (\textbf{86.8}) & {96.8}\\
        \bottomrule
    \end{tabular}
    \end{adjustbox}
    \vspace{-5mm}
    \label{tab:workload}
\end{table}
We further use a simple yet effective preprocessing technique: zeroing out all presynaptic neurons that have a low firing activity to further improve the number of silent neurons. We take the trained SNN and mask the neurons with only one output spike throughout all timesteps. We find that with a very small number of fine-tuning ($<$5 epochs), the accuracy can be fully recovered, as shown in Figure~\ref{fig:exp:fine-tune}. {Please note that this preprocessing technique aims to maintain the accuracy of the original workload instead of improving it.}
During hardware execution, the compressor will discard the output neurons that have 0 or only 1 output spike. From Table \ref{tab:workload}, we see that preprocessing effectively creates up to $1.1\times$ more silent neurons\footnote{The source codes can be found at \url{https://github.com/RuokaiYin/LoAS}}.

\minisection{Hardware Configuration: } We evaluate \name with the configuration in Table~\ref{tb:hw_config}. In our experiments, we configure the \name to support SNNs running with 4 timesteps. We use 16 TPPEs, each with 5 accumulators (1 12-bit pseudo-accumulator and 4 10-bit correction accumulators) and 1 inner-join unit. Inside each inner-join unit, there is 1 fast prefix-sum circuit and 1 laggy prefix-sum circuit. The fast prefix-sum circuit can generate the offsets in a single cycle. The laggy prefix-sum circuit contains 16 adders and a 128-bit buffer. It generates the offset results in 8 cycles. The TPPE also has 2 depth-8 FIFOs (for correction purposes) and 2 128-bit buffers (for holding bitmasks). Finally, a 128-byte buffer is equipped inside the TPPE to hold the non-zero weights from $fiber-B$.
We allocate 256 KB (double-buffered) for the global cache. For our workloads, this memory size is enough to capture good on-chip data reuse and keep all TPPEs busy.

\minisection{Baseline: }
As discussed previously, there are currently very limited spMspM accelerators available for dual-sparse SNNs. 
As a result, we construct our baselines in the following way,
{We first pick three popular ANN spMspM accelerators that use \textbf{\texttt{IP}}, \textbf{\texttt{OP}}, and \textbf{\texttt{Gust}} dataflow: SparTen~\cite{sparten}, GoSPA~\cite{gospa}, and Gamma~\cite{gamma}. }
We then envision that a dual-sparse SNN (with 4 timesteps and 8-bit weights) is naively running (sequentially processing its timesteps) on these accelerators. 
\begin{table}[t]
\centering
\caption{Configuration of the \name System.}
\vspace{-3mm}
\begin{adjustbox}{max width =0.75\linewidth}
\begin{tabular}{l | l}
\toprule
TPPEs & 16 TPPEs, 8-bit weight\\
Inner-join unit & 16 Inner-join units\\
Global cache & 256 KB, 16 banks, 16-way associative \\
Crossbars & $16\times16$ and $16\times16$, swizzle-switch based \\
Main memory & 128 GB/s over 16 64-bit HBM channels \\
\bottomrule
\end{tabular}
\end{adjustbox}
\vspace{-5mm}
\label{tb:hw_config}
\end{table}
{To be conservative, we place the $t$ dimension at the innermost loop of the original \textbf{\texttt{IP}}, \textbf{\texttt{OP}}, and \textbf{\texttt{Gust}} dataflow.\footnote{Adding the $t$ dimension anywhere else will bring more data traffic, thus worsening the performance.} 
We then make essential simplifications for the two accelerators. 
For example, we remove the multipliers in these designs. 
To make a fair comparison, we configure all designs to have 16 PEs and the same global SRAM size. We call these three baselines SparTen-SNN, GoSPA-SNN, and Gamma-SNN.}

We implement the key components of \name and our hardware baselines in RTL and synthesize them using Synopsys DC compiler at 800MHz with 32 nm technology. A 128 GB/s High-Bandwidth Memory (HBM) module is connected to \name as the off-chip memory. We use CACTI~7.0~\cite{cacti} to model the memory components. We built a simulator in Python to model the cycle-level behavior of \name and the baselines by tiling the loop and mapping it to hardware.

\section{Experimental Results}
\label{sec:eval_result}

\subsection{Hardware Evaluation}
\begin{figure}[t]
\centering
{\hspace{-4mm}\includegraphics[width=0.8\linewidth]{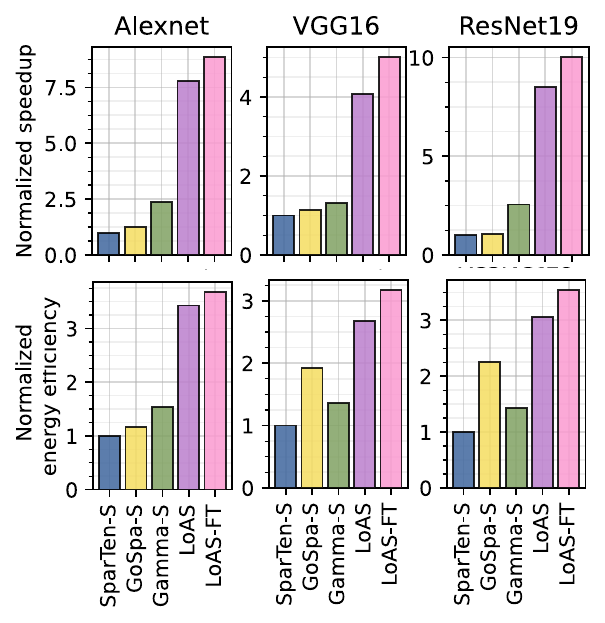}}
\vspace{-4mm}
  \caption{Performance and efficiency comparison between SparTen-SNN, GoSPA-SNN, {Gamma-SNN}, and \name (with and without fine-tuned (FT) pre-processed) architectures across three SNN workloads. All numbers are normalized to that of the SparTen-SNN baseline.}
  \label{fig:exp:perf-main}
   \vspace{-5mm}
\end{figure}

\minisection{Overall Performances: }Figure~\ref{fig:exp:perf-main} compares the performance between {three dual-sparse SNN accelerator baselines (SparTen-SNN, GoSPA-SNN, and Gamma-SNN)} and \name (with and without fine-tuned preprocessing) on three SNNs (speedup w.r.t the cycle numbers of the SparTen-SNN).

The first observation is that \name significantly outperforms the other {three accelerator baselines in all cases, obtaining average speed-ups of $6.79\times$ (vs. SparTen-SNN), $5.99\times$ (vs. GoSPA-SNN), and $3.25\times$ (vs. Gamma-SNN)}. This is due to \name leverages \textbf{\texttt{FTP}} dataflow. The \textbf{\texttt{FTP}} dataflow completely unleashes \name from the intra-PE latency penalty of sequentially running the timesteps. It also enables \name to invoke less on-chip and off-chip data communications across timesteps.
The second observation is that \name's performance gain is highly correlated with the sparsity of matrix A. This relationship is expected since our workloads are extremely sparse on matrix B; thus, the overall computation is matrix-A-bounded. Consequentially, the performance of two baselines suffers more from sequentially running timesteps through matrix A with less sparsity. However, \name will not get this sequentially running penalty. As a result, \name achieves from $4.08\times$ speedup (vs. SparTen-SNN) on VGG16 (highest matrix A sparsity) to $8.51\times$ speedup (vs. SparTen-SNN) on ResNet19 (lowest matrix A sparsity). Finally, we observe that with the help of pre-processing (removing the neurons that only spike one time), \name further improves the performance by $20\%$ on average. This is because the pre-processing technique helps to increase the density of silent neurons (Section~\ref{sec:mem_compress}), which \name is able to completely avoid the data communications and computations. Figure~\ref{fig:exp:perf-main} also compares the energy efficiency of \name and {three baselines on different SNN workloads. It is observed that \name (with preprocessing) achieves ($3.68\times$, $3.09\times$, {$2.40\times$}), ($3.17\times$, $1.50\times$, {$2.33\times$}), and ($3.54\times$, $1.34\times$, {$2.47\times$}) higher energy efficiency over (SparTen-SNN, GoSPA-SNN, and Gamma-SNN}) on Alexnet, VGG16, and ResNet19.

\begin{figure}[t]
\centering
{\hspace{-5mm}\includegraphics[width=0.9\linewidth]{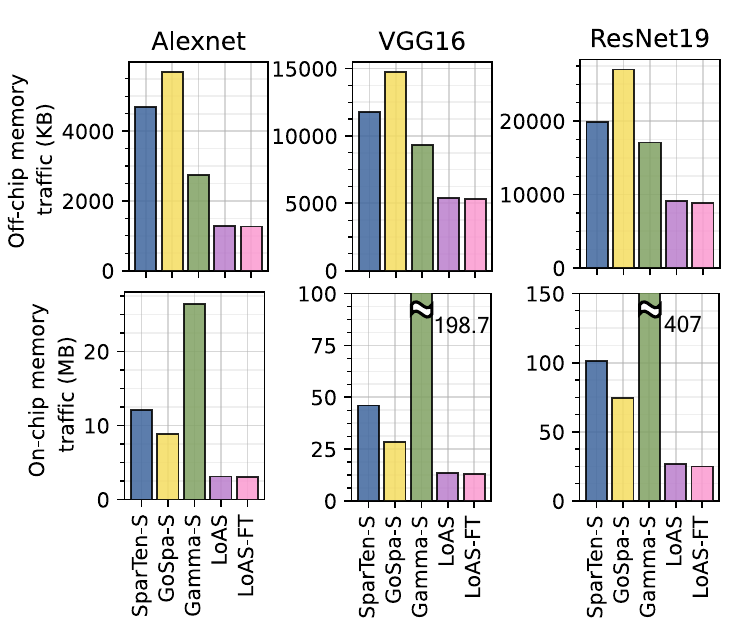}}
    \vspace{-5mm}
  \caption{Off-chip traffic (KB) and on-chip memory traffic (MB) for SparTen-SNN, GoSPA-SNN, {Gamma-SNN}, and \name (with and without pre-processed) architectures across three SNN workloads.}
  \label{fig:exp:mem-main}
   \vspace{-3mm}
\end{figure}
\minisection{Detailed Analysis:} We next explain the performance gains of \name. Owing to the \textbf{\texttt{FTP}} dataflow, \name has much less on-chip and off-chip memory traffic than the two baselines. As shown in Figure~\ref{fig:exp:mem-main}, compared to SparTen-SNN (\textbf{\texttt{IP}}), \name has $3.93\times$($3.70\times$), $3.57\times$($2.22\times$), and $4.07\times$($2.24\times$) less on-chip SRAM (off-chip DRAM) access on Alexnet, VGG16, and ResNet19, respectively. 
This behavior is expected since \textbf{\texttt{IP}} dataflow design like SparTen is known for having poor input data reuse. This inefficient input data reuse pattern is exacerbated by the extra temporal dimension ($t$-dim) in SNN workloads. While \textbf{\texttt{FTP}} dataflow is a variant of inner-product, it does not incur any extra executions on the $t$-dim since it parallelizes the $t$-dim at the inner-most loop.

\begin{figure}[t]
\centering
{\hspace{-5mm}\includegraphics[width=0.9\linewidth]{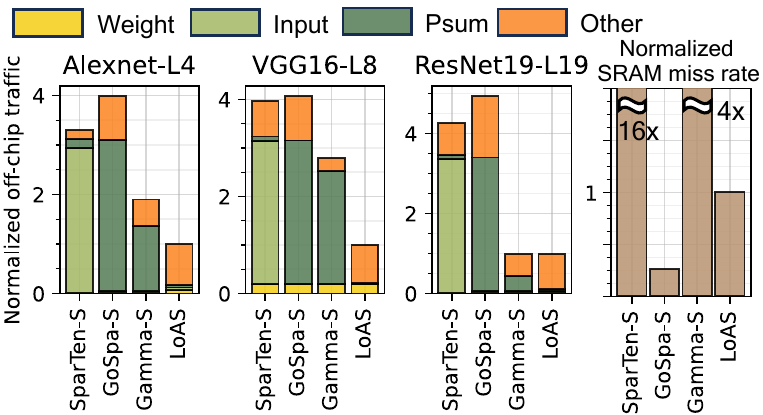}}
\vspace{-3mm}
  \caption{Normalized Off-chip traffic with breakup for SparTen-SNN, GoSPA-SNN, { Gamma-SNN}, and \name (with pre-processed) architectures across three SNN layer workloads. The normalized SRAM cache miss rate is also provided for the ResNet19 layer workload. All numbers are normalized to that of \name.}
  \label{fig:exp:layer-breakup}
   \vspace{-3mm}
\end{figure}

Not surprisingly, compared to GoSPA-SNN (\textbf{\texttt{OP}}), \name still achieves $2.87\times$($4.49\times$), $2.19\times$($2.78\times$), and $2.98\times$($3.03\times$) less on-chip SRAM (off-chip DRAM) access on Alexnet, VGG16, and ResNet19, respectively.
This behavior is also expected even though \textbf{\texttt{OP}} dataflow design is known to have excellent input data reuse (on average, GoSPA-SNN has $1.45\times$ less SRAM traffic than SparTen-SNN). The inefficiency for GoSPA-SNN comes from the partial sum (\texttt{psum}) matrices. Because of the extra $t$-dim in SNNs, the size of \texttt{psum} matrices expands with the number of timesteps. GoSPA's design allocates a small on-chip memory for the \texttt{psum}. The \texttt{psum} matrices that cannot fit on-chip must be written to off-chip DRAM and read back later for reduction. This incurs significant off-chip memory traffic.

{Finally, compared to Gamma-SNN (Gust), LoAS is able to achieve $2.16\times$, $1.76\times$, and $1.91\times$ less DRAM accesses. This result is aligned with Gust dataflow's ability to reduce off-chip partial row accesses through on-chip SRAM and mergers. While reducing the DRAM accesses, Gamma's SRAM accesses are exacerbated by the $t$-dim in SNNs. This ends up with on average $13.4\times$ more SRAM traffic than \name.}

To better visualize the aforementioned analysis, we provide a memory traffic breakup in Figure~\ref{fig:exp:layer-breakup} for the three SNN layers in Table~\ref{tab:workload}. As shown in the figure, SparTen-SNN has the largest input off-chip traffic, and GoSPA-SNN has the largest \texttt{psum} off-chip traffic across all workloads. {Among the three baselines, Gamma-SNN has the smallest off-chip traffic footprint due to Gust dataflow's on-chip reuse of partial rows.} GoSPA-SNN has the largest off-chip traffic for compressed format due to its CSR format for each spike. We notice that \name has slightly larger ($2.1\times$) off-chip traffic for compressed format compared to SparTen-SNN. This is because we need extra bitmasks to mark the position of non-silent neurons, while in SparTen-SNN, we can directly leverage the input spike trains. Nevertheless, this overhead is negligible compared to \name's saving on off-chip traffic for other quantities. Figure~\ref{fig:exp:layer-breakup} also provides the normalized SRAM cache miss rate for the layer workload in ResNet19. SparTen-SNN has a $16\times$ higher miss rate(1.47\%) compared to \name. GoSPA-SNN has the lowest miss rate due to its Output-stationary dataflow. However, the tradeoff is the higher off-chip traffic of \texttt{psums}. {Gamma-SNN has a higher SRAM miss rate than GoSPA-SNN and \name. The reason is that the extra $t$-dim enlarges the partial row traffic by $t$ times. Some of the extra traffic cannot be held in the on-chip SRAM, thus leading to the cache eviction.} Overall, the cache miss rate results align with the off-chip traffic trends. Since we set all the baselines to have the same global cache size, the reduction in the memory traffic reflects \name's improvement in both speedup and energy efficiency.

\begin{table}[t]
\centering
\caption{Area and Power breakdown of \name (left) and one TPPE (right).}
\vspace{-3mm}
\begin{adjustbox}{max width =0.9\linewidth}
\begin{tabular}{lcc|lcc}
\toprule
\textbf{Components} & \textbf{Area (mm\textsuperscript{2})} & \textbf{Power (mW)}& \textbf{TPPE units} & \textbf{Area} & \textbf{Power}\\
\midrule
16 TPPEs & 0.96 & 45.1 & Accumulators& 2e-3 & 0.16\\
16 PLIFs & 0.02 & 1.2 & Fast Prefix & 0.04 & 1.46 \\
Global cache & 0.80 & 124.5 & Laggy Prefix & 5e-3& 0.32\\
Others & 0.30 & 18.1 & Others & 0.01 & 0.88\\
\midrule
\textbf{Total} & \textbf{2.08} & \textbf{188.9} & \textbf{TPPE total} & \textbf{0.06} & \textbf{2.82} \\
\bottomrule
\end{tabular}
\end{adjustbox}
\label{table:area_power}
\end{table}

\begin{figure}[t]
\centering
\includegraphics[width=0.8\linewidth]{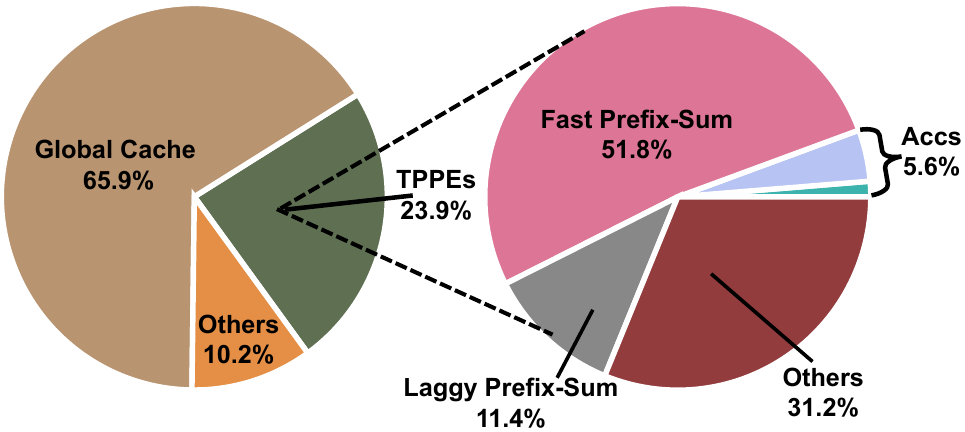}
\vspace{-3mm}
  \caption{On-chip power breakup of \name. Accs stands for the accumulators, which include 1 pseudo-accumulator and 4 correction-accumulators.}
  \label{fig:exp:power-breakup}
   \vspace{-5mm}
\end{figure}

\minisection{Area and Power:}Table~\ref{table:area_power} shows the area and power breakdown of \name with the configuration in Table~\ref{tb:hw_config}. Inside each TPPE, one single fast prefix-sum circuit dominates both the area (66.7\%) and power (51.8\%).
Original SparTen\cite{sparten} even requires two fast prefix-sum circuits for both inputs and weights.\footnote{This is not the case in SparTen-SNN. Since the input spikes are bitmasks and data at the same time, thus SparTen-SNN only requires one fast prefix-sum circuit.}
Thanks to the laggy prefix-sum circuits (8.3\% of area and 11.4\% of power) we proposed, \name only requires one fast prefix-sum circuit inside each TPPE. At the system level, the global SRAM cache dominates both the power and area, which aligns with the previous works~\cite{spada,flexagon, gamma}. Figure~\ref{fig:exp:power-breakup} provides a visualization of the power breakup.

\begin{figure}[t]
\centering
\hspace{-5mm}\includegraphics[width=\linewidth]{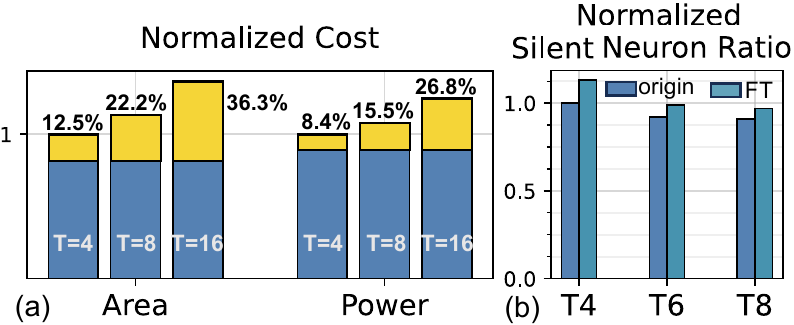}
\vspace{-3mm}
  \caption{(a) The scalability of TPPE with increasing timesteps. The yellow region denotes the portion that grows with the timesteps. {(b) The scalability of the ratio of silent neurons (sparsity of matrix A) with increasing timesteps. All values are normalized to the original silent neuron ratio at the timestep of 4.}}
  \label{fig:exp:scale_study}
   \vspace{-5mm}
\end{figure}
\subsection{Ablation Studies}
\minisection{Temporal Scalability Studies: }In our experimental settings, we configured the TPPE inside \name to run the SNNs with 4 timesteps. Most state-of-the-art SNN algorithms~\cite{deng2022temporal,fang2021deep} usually use a timestep equal to or less than 8. So, we want to understand how TPPE scales with the timesteps. 
Figure~\ref{fig:exp:scale_study}(a) shows that TPPE scales well with the timesteps. The reason is that all TPPE components other than accumulators and the input data buffer are agnostic to the number of timesteps. Even at 16 timesteps, the TPPE only increases its area (power) by 1.37$\times$ (1.25$\times$) compared to 4 timesteps. {We also showcase how the ratio of silent neurons in VGG16 scales with the number of timesteps. Figure~\ref{fig:exp:scale_study}(b) shows that with the help of the pre-processing technique, even at the timestep of 8, we can still have a similar ratio of silent neurons as the timestep of 4. However, it is very likely to have fewer silent neurons when we have even larger timesteps ($>8$). This is one of the challenge that LoAS needs to face when scaling up on the number of timesteps.}

\begin{figure}[t]
\centering
\includegraphics[width=0.8\linewidth]{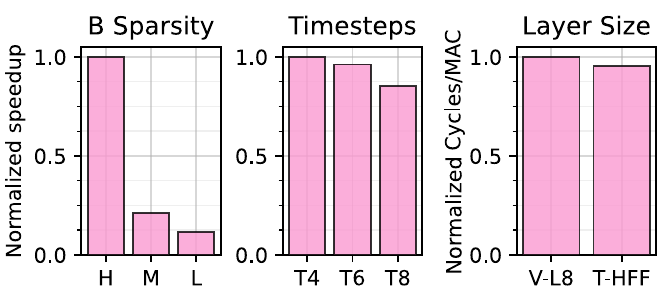}
\vspace{-3mm}
  \caption{{Scalability of \name across different sparsity patterns of matrix B, number of timesteps, and layer size.}}
  \label{fig:exp:scale_perf}
   \vspace{-3mm}
\end{figure}

{\minisection{Scalability Study: }
Figure~\ref{fig:exp:scale_perf} further shows how the overall performance of \name scales with different quantities. We first test \name running on VGG16 with average sparsity of B (weight) at $98.2\%$(High), $68.4$(Medium), and $25\%$(Low). The result shows that \name's performance is highly sensitive to the sparsity level of B. When we scale the sparsity from $98.2\%$ to $25\%$, the performance scales down by roughly $88\%$. We also find that \name's performance scales pretty well on timesteps. \name only loses roughly $14\%$ of performance when increasing the number of timesteps by $2\times$. Finally, we test \name's scalability on layer size. We compare one layer from VGG16 and the hidden feed-forward (HFF) layer from SpikeTransformer~\cite{yao2024spike}. The results show that \name scales pretty well, even on the layer with a larger parameter size.}

\begin{figure}[t]
\centering
\includegraphics[width=0.75\linewidth]{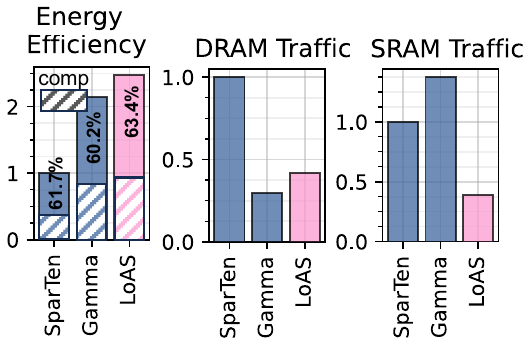}
\vspace{-5mm}
  \caption{{Normalized energy efficiency and memory traffic between SNNs (\name., T=4) vs. ANN baselines (SparTen, Gamma).}}
  \label{fig:exp:ann_snn}
   \vspace{-3mm}
\end{figure}

\minisection{Dual-sparse SNN vs. Dual-sparse ANN: }
% \lipsum[3-4]
In this work, we focus on providing insights for the community on how the spMspM acceleration works on SNNs. However, it is unavoidable to discuss the comparison between SNNs and ANNs.
In Figure~\ref{fig:exp:ann_snn}, we show the comparison of normalized energy efficiency and memory traffic between SNNs (\name) and ANNs (SparTen~\cite{sparten}) { and Gamma~\cite{gamma}} running VGG16 workload.
We use the VGG16 workload in Table~\ref{tab:workload} for \name. ANN-version of VGG16 has 8-bit weights ($98.2\%$ sparsity) and activations ($43.9\%$ sparsity). Overall, the SNN running on \name has roughly $2.5\times$ { and $1.2\times$} energy efficiency compared to the ANNs { running on SparTen and Gamma, respectively}. We observe that around $60\%$ of energy contributes to the data movement for both networks. We, therefore, also include the DRAM and SRAM traffic comparison in Figure~\ref{fig:exp:ann_snn}. It shows that SNNs, on average, have $\sim60\%$ less memory traffic compared to SparTen-ANN. The less memory traffic comes from less input bitwidth (4-bit vs. 8-bit) and higher input sparsity ($79.6\%$ vs. $43.9\%$), thanks to SNN's features of unary activation and sparse spike activity (\ref{sec:bg:snn_con}). { Not surprisingly, Gamma-ANN has lower overall DRAM accesses compared to LoAS due to its Gust dataflow~\cite{gamma}. The tradeoff is $3.5\times$ more SRAM traffic, which explains why the LoAS has a slightly higher overall energy efficiency.}

\begin{figure}[t]
\centering
\includegraphics[width=0.9\linewidth]{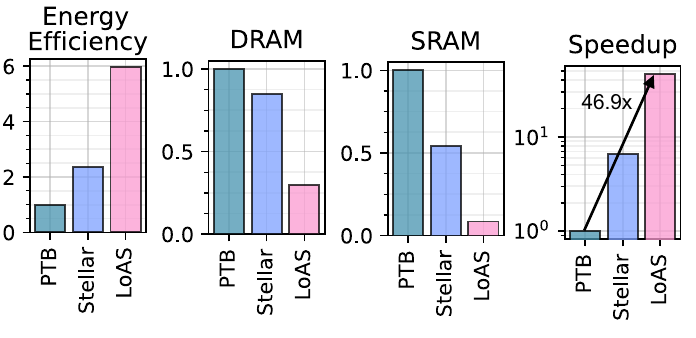}
\vspace{-5mm}
  \caption{{ Normalized performance comparison between dual-sparse SNN accelerator (\name) vs. dense SNN accelerator baselines (PTB, Stellar).}}
  \label{fig:exp:ptb_loas}
  \vspace{-2mm}
\end{figure}

\minisection{Dual-sparse SNN vs. Dense SNN: }
To show the benefits of dual-sparsity in SNNs, we compare \name with the prior { dense SNN systolic-array accelerators, PTB~\cite{ptb} and Stellar~\cite{mao2024stellar}}, running dense VGG16 with 4 timesteps. For a fair comparison, we set the array size for PTB to be $16\times4$, which generates 16 full-sum outputs for 4 timesteps in parallel (same as \name). { We further configure Stellar to the same array size.} We leverage ScaleSim~\cite{scalesim} to estimate both baselines' memory traffic and cycle counts. We show the comparison in Figure~\ref{fig:exp:ptb_loas}. We first observe that \name has roughly $6\times$ higher energy efficiency compared to PTB, mainly resulting from the $3\times$ ($12.5\times$) less DRAM (SRAM) traffic. { Compared to Stellar, \name has roughly $2.5\times$ higher energy efficiency, as well as the $2.7\times$ ($6.6\times$) less DRAM (SRAM) traffic.} We also observe that \name has $46.9\times$ speedup against PTB. This is primarily due to the data sparsity and the difference between PTB's partially temporal parallel (Section~\ref{sec:bg:prior_snn}) and \name's fully temporal parallel mechanism. { We observe that Stellar outperforms PTB across all matrices. This is mainly due to Stellar's optimized spatiotemporal row-stationary dataflow and its spike-skipping technique. However, compared to Stellar, we are still able to achieve roughly $7.1\times$ speedup due to LoAS's capability to leverage the dual-sparsity.} Please note that we do not compare with the SpinalFlow~\cite{spinalflow} due to its temporal encoding achieves limited accuracy on challenging learning tasks~\cite{ptb,comsa2020temporal}.

\section{Related Work}
\label{sec:discussion}
Except for the prior SNN dense accelerator works we discussed in Section~\ref{sec:bg:prior_snn}, there also exists prior works that try to leverage the sparsity in SNNs. In~\cite{chen202167}, a neuron filter unit is leveraged to only fetch the weight if there is a 1-spike. However, dual-sparsity (both spike and weight sparsity) is not considered. In~\cite{chen2022cerebron}, the dual-sparsity of SNN is considered to skip the unmatched computation. However, the weights and spikes are fetched in a dense format without any compression from the off-chip memory, thus failing to save data movement costs. In this work, \name leverages the dual-sparsity in SNNs from both computation and data movement.

As we discussed, PTB processes the timesteps in a partially parallel manner. Even if one re-configures the PTB to run all timesteps in parallel (time-window=1), it still differs from \name in the loop ordering. 
{In PTB's loop ordering, $t$-dim is placed between $m$-dim and $n$-dim, while \name places the $t$-dim in the inner-most loop. 
As discussed in Section~\ref{sec:dataflow}, \name's loop ordering brings more efficiency in spMspM operation. Moreover, PTB targets accelerating workloads with time-series data from DVS sensors~\cite{dvscifar10}, where the timestep is usually large ($>100$). On our workloads, where the timesteps are small ($<8$), PTB experiences low hardware utilization.}
In~\cite{liu2022sato}, processing timesteps in parallel is also studied. However, they target the temporal-coded SNN workloads, and the loop ordering is not discussed. { Finally, as discussed in Section~\ref{sec:bg:prior_snn}, Stellar~\cite{mao2024stellar} is another work that also tries to process timesteps in parallel. However, it targets the non-LIF, FS-coded SNNs and does not support the dual-sparsity.}

\section{Conclusion}
\label{sec:conclusion}
In this work, we observe that naively running dual-sparse SNNs on existing spMspM accelerators exhibits sub-optimal efficiency due to the latency and memory traffic penalty brought by processing timesteps.
To improve the efficiency, we propose a fully temporal-parallel dataflow (FTP), which avoids the above problems.
To maximize the benefits of FTP, we propose FTP-friendly spike compression and inner-join mechanism.
We also build \name, a novel architecture that exemplifies the FTP dataflow.
With the help of both FTP-friendly compression and inner-join, \name demonstrates significant speedup (up to $8.51\times$) and energy reduction (up to $3.68\times$) compared to prior dual-sparse accelerator baselines.

%%%%%%% -- PAPER CONTENT ENDS -- %%%%%%%%

%%%%%%%%% -- BIB STYLE AND FILE -- %%%%%%%%
\bibliographystyle{IEEEtranS}
\bibliography{refs}
%%%%%%%%%%%%%%%%%%%%%%%%%%%%%%%%%%%%

\end{document}